\documentclass[conference]{IEEEtran}
\IEEEoverridecommandlockouts
\usepackage{cite}
\usepackage{amsmath,amssymb,amsfonts}
\usepackage{algorithmic}
\usepackage{graphicx}
\usepackage{textcomp}
\def\BibTeX{{\rm B\kern-.05em{\sc i\kern-.025em b}\kern-.08em
    T\kern-.1667em\lower.7ex\hbox{E}\kern-.125emX}}
\begin{document}

\title{A Comparative Study on using Principle Component Analysis with Different Text Classifiers}

\author{\IEEEauthorblockN{Ahmed I. Taloba}
\IEEEauthorblockA{\textit{Information System Department} \\
\textit{Faculty of Computers and Information}\\
\textit{Assiut University}\\
Assiut, Egypt \\
Taloba@aun.edu.eg}
\and
\IEEEauthorblockN{D. A. Eisa}
\IEEEauthorblockA{\textit{Mathematics Department} \\
\textit{Faculty of Science}\\
\textit{Assiut University}\\
New Valley, Egypt \\
dalia\_ah@yahoo.com}
\and
\IEEEauthorblockN{Safaa S. I. Ismail}
\IEEEauthorblockA{\textit{Mathematics Department} \\
\textit{Faculty of Science}\\
\textit{Assiut University}\\
New Valley, Egypt \\
safaasobhy1982@scinv.au.edu.eg}
}

\maketitle

\begin{abstract}
Text categorization (TC) is the task of automatically organizing a set of documents into a set of pre-defined categories. Over the last few years, increased attention has been paid to the use of documents in digital form and this makes text categorization becomes a challenging issue. The most significant problem of text categorization is its huge number of features. Most of these features are redundant, noisy and irrelevant that cause over fitting with most of the classifiers. Hence, feature extraction is an important step to improve the overall accuracy and the performance of the text classifiers. In this paper, we will provide an overview of using principle component analysis (PCA) as a feature extraction with various classifiers. It was observed that the performance rate of the classifiers after using PCA to reduce the dimension of data improved. Experiments are conducted on three UCI data sets, Classic03, CNAE-9 and DBWorld e-mails. We compare the classification performance results of using PCA with popular and well-known text classifiers. Results show that using PCA encouragingly enhances classification performance on most of the classifiers.
\end{abstract}


\section{Introduction}
Most text analysis, such as text categorization (TC), includes an essential step of feature extraction to find the best set of features that assimilate each text. Text categorization is one of the central problems in text mining and information retrieval, where it is the task of classifying documents by the words of which the documents include. Several machine learning algorithms have been developed for text classification, e.g.:  decision tree (J-48) \cite{BIB:R91}, k-nearest neighbor (KNN) \cite{BIB:R93}, support vector machine(SVM) \cite{BIB:R94} and random forests (RF) \cite{BIB:R95}. Thus, these text classifiers give acceptable accuracy with high dimensional data such as text.\\\\
 There are many applications of text categorization such as topic detection \cite{BIB:R1}, phishing – email detection\cite{BIB:R2},  author identification \cite{BIB:R3}and etc.\\\\

In text categorization, a text or a document is  always represented as a bag of words. The high dimensionality of the feature space emerged as a critical problem due to this representation. This huge number of features in the feature vector results in time complexity \cite{BIB:R4} and poor performance of the classifier, so the number of input variables have to be reduced before applying a text categorization algorithm. The reduction of the feature space makes the training faster, improves the accuracy of the classifier by removing the noisy features and avoid overfitting.\\\\
The dimensionality reduction in text categorization can be made in two different ways: feature selection and feature extraction. In feature selection techniques,  the most relevant variables are kept from the original data set, where as in feature extraction  techniques, the original vector space is transformed into a new one with some special characteristics and the reduction is made in a new vector space. From the most popular feature extraction techniques are principle component analysis (PCA) \cite{BIB:R5}\cite{BIB:R6}, latent semantic indexing\cite{BIB:R7}, clustering methods\cite{BIB:R8} and etc.\\\\
Among these many methods, PCA has attracted a lot of attention. PCA is a tool to reduce feature vector to lower dimension while retaining most of the information. It has been used since the early 90s in text processing tasks \cite{BIB:R8}\cite{BIB:R9}. PCA’s key advantages are its low noise sensitivity, reduce the need for capacity and memory, does not need large computations and increased efficiency given the classifiers taking place in a smaller dimensions \cite{BIB:R18}.\\\\
In the current study, first, the documents are processed with the following steps; initially the documents are collected, followed by pre-processing, indexing, feature extraction, classification algorithms and performance measure. After the documents are processed, the documents are converted to a huge matrix which we have to reduce and that we do after indexing, where we use PCA as a feature extraction. However, PCA obtain a good feature subset in a less time cost. Then, after feature extraction step, the extracted features will be passed to different classifiers such as random forest (RF), support vector machine (SVM), decision tree (J-48) and k-nearest neighbours (KNN). Finally, the effectiveness of each classifier is computed; however, sensitivity, specificity and accuracy are mostly used. Experiments are conducted on three UCL data sets \cite{BIB:R23}, which are Classic03, CNAE-9 and DBWorld e-mails collection for text classification.

\section{Previous Works}
In \cite{BIB:R14} they try to speed feature extraction by using a method that folds together Unicode conversion, forced lowercasing, word boundary detection, and string hash computation. The method they found require less computation and less memory to find the integer hash features result in classifiers with equivalent statistical performance to those built using string word features.\\\\
In \cite{BIB:R15} they modulate the classifier performance. They use two-stage feature selection and feature extraction methods. In the first stage, according to the importance for classification each term within the document is ranked using the information gain (IG) method. In the second stage, the dimension reduction step, they use genetic algorithm and principle component analysis for feature selection and feature extraction after ranking the terms in descending order of importance.They apply their model on two UCI data sets, Reuters-21,578 and Classic03 which are collected for text categorization. The experimental results show that their proposed model actualizes high classification effectiveness as measured by precision, recall and F-measure.\\\\
In \cite{BIB:R16} they try to solve the high dimension of the feature vector problem for text categorization. They use a multistage model to enhance the overall accuracy and the performance of classification. In the first stage, the documents are processed and each document is represented by a bag of words. In the second step each term within the documents are ranked according to their importance for classification using the information gain (IG). Then the third stage is the attribute reduction step based on rough set which is carried out on the terms which are ranked according to their importance. Finally the extracted features are then passed to naive bayes and KNN classifier. They apply their model on three UCI data sets, Reuters-21578, Classic04 and Newsgroup 20.\\\\
In \cite{BIB:R2} they try to solve a critical text classification application, phishing email detection. They use two feature selection techniques - chi-square, information gain ratio and two feature extraction techniques – principal component analysis, latent semantic analysis are used for extracting the features that improve the classification accuracy. The data set used is prepared by collecting a group of e-mails from the well known publicly available corpus that most authors in this area have used. Phishing data set consisting
1,000 phishing emails received from November 2004 to August 2007 provided by Monkey web site and, 1,700 Ham email from Spam Assassin project.\\\\
In \cite{BIB:R117} a term frequency (TF) with stemmer-based feature extraction method is proposed for document classification. The classification accuracy was calculated using J-48 classification algorithm. The effectiveness of proposed method was investigated and compared
against well known other feature extraction techniques.
\section{The Proposed Method}
The stages in text categorization (TC) we will follow in our work is composed of the following steps:
\subsection{Documents Collection}
This is the initial  step of any text categorization process in which documents of several format like .html, .pdf, .doc, web content and etc. are collected.
\subsection{Preparing the Text for Classification}
Pre-processing of documents is an essential task during text categorization process before using the classifier in order to transform documents, which typically are strings of characters, into a set of words, which is a suitable representation for the learning algorithm, and at the same time enriching their semantic meaning. Documents filtering and stemming are applied in the extracted bag of words we have to reduce the dimensionally.

\subsubsection{Stop Words Removal}
Words such as pro-nouns, prepositions and articles, etc. in texts does not affect on the meaning of the documents. These words are called stop words. The removing of these words from the bag of words is an essential step such that these words are useless for purposes of retrieval. The stop words are not measured as a keywords in text mining applications \cite{BIB:R13}. Example for stop words: "the","a", "an", "with", etc. Removing stop words reduces the term space such that these words make the text look heavier and less important for analysts.
\subsubsection{Stemming}
This step is used to identify the words to its root. For example, extracted, extracting, extracts, extraction all can be stemmed to the word "extract" \cite{BIB:R11}. The aim of this step is to eliminate various suffixes, to decrease the number of words, to have accurately matching stems, to save time and memory space.
\subsection{Indexing}
After the pre-processing step, each document is represented by a bag of words. We have to find the technique that create a vector representation of each document. So this step decrease the complexity of the documents and make them easier to handle. Each word is assigned a weight based on its number of times it appears in the document as shown in the following matrix. This process is known as term weighting.\\\\
$\left(
  \begin{array}{ccccc}
    T_{1} & T_{2} & $..$ & T_{at} & C_{i} \\
  D_{1} & W_{11} & $..$ & W_{t1} & C_{1} \\
  D_{2} & W_{12} & $..$ & W_{t2} & C_{2} \\
  . & .& $..$ & $.$ & .\\
  . & .& $..$ & $.$ & .\\
  D_{n} & W_{1n} & $..$ & W_{tn} & C_{n} \\
  \end{array}
\right)$\\\\
\\\\
Where $W_{in}$ is the weight of word \emph{i} in the document \emph{n}. We have to know that there are several ways of calculating weight, such as boolean weighting, word frequency weighting, TF-IDF, entropy, etc. A commonly used term weighting method is the so-called TF-IDF (term frequency - inverse document frequency) weighting, which is a numerical statistic measure used to reflect how important a word is to a document in a certain class of collection or corpus. The importance increases proportionally to the number of times a word appears in the document but is offset by the frequency of the word in the corpus. For calculating the TF-IDF weight of a term in a particular document, it is necessary to calculate: term frequency (TF(t,d)) which is the number that the word \emph{t} occurred in the document \emph{d}, document frequency (DF(t)) is number of documents in which the term \emph{t} occur at least once and inverse document frequency (IDF) that can be calculated from document frequency using the following formula

\begin{equation}\label{e4}
  IDF = \log{(\frac{\tiny {num \enspace of \enspace documents}}{{\tiny num \enspace of \enspace documents \enspace with \enspace word \enspace i}})}
\end{equation}

The inverse document frequency of a word is low if it occurs in many documents and is high if the word occurs in only few documents.
The measure of word important can be calculated by using the product of the term frequency and the inverse document frequency (TF * IDF).

\subsection{Feature Extraction}
After the preprocessing and indexing step we have a matrix of high dimension. Having too many features cause many problems such as overfitting, reduce the accuracy of the classifier and cause high time complexity. Feature extraction will reduce the matrix size and this will improve the scalability, efficiency and accuracy of the classifiers. The main idea of feature extraction is to reduce the features dimension by creating new combinations of attributes (words). Choosing right features and deciding how to encode them to be an input for the classifier can have an enormous impact on the classifier ability to extract a good model.\\

Principle component analysis (PCA) is one of the most popular statistical technique for feature extraction.PCA help to improve the discriminative power of the classifiers. PCA is a useful statistical technique that many applications can be used it, such as face recognition and image compression, and is a popular technique for finding patterns in data of high dimension without losing important information such as text categorization. It is used to project the original feature space onto a lower dimensional subspace. The key idea in PCA is to find a subset of variables from a larger set, based on which original variables have the highest correlations with what is called principal components \cite{BIB:R18}. The number of the principle components is less than or equal to the number of the original features. It is an acceptable choice for reducing the dimensionality of highly dimensional data.

\subsection{Text Classifiers}
There are many classifiers that have been developed for variety of tasks in text classification and they give acceptable accuracy. Among them we will use the following classifiers and show how the accuracy improved after using PCA as a feature extraction:\\

\begin{itemize}
\item \emph{Random Forest (RF):}
  RF is a very good, powerful, robust and versatile learning technique, however it is a promise choice for high-dimensional text data. It is introduced in 2000s \cite{BIB:R95}, it is a popular classification method which builds multiple decision trees (not only one), which are used to determine the final outcome. For classification problems, the ensemble of simple trees vote for the most popular class. One of the most known forest construction procedures, proposed by Breiman, is  a subspace of features which are chosen randomly at each node to grow branches of the decision trees, then bagging method is used to generate
training data subsets for building individual trees, finally combination of all individual trees are formed to form random forests
model \cite{BIB:R95}.\\

  \item \emph{Support Vector Machine (SVM):}
  SVM has been recognized as one of the most effective text categorization method. It gives high classification accuracy especially in highly dimensional data such that it controls complexity and overfitting issued. The time taken for each process is less than the other classifiers. So that it becomes an acceptable choice for large data set as textual data. SVMs are designed to handle high-dimensional data. SVM was developed in 1995 by Cortes and Vapnik \cite{BIB:R17}. Its core idea behind SVM is to find an optimal hyper plane between sets of hyper plane that maximize hyper plane margin, which is the distance from hyper plane to the nearest point of the pattern \cite{BIB:R17} \cite{BIB:R19}. The document representatives which are closest to the decision surface are called the support vectors. SVM is primarily used to maximize the margin, which will ensure that the input pattern would be classified correctly \cite{BIB:R20}.\\
  The aim of SVM is to find out the best possible classification function in order to differentiate between members of two classes in the training data in a two-class learning task.\\

    \item \emph{Decision Tree Algorithm (J-48):} The decision tree reconstructs the manual categorization of training documents by making well defined true/false-queries in the form of a tree structure. In the decision tree structure, leaves represent the corresponding
category of documents and branches represent conjunctions of features that lead to those categories. The tree expands until each and every text is categorized correctly or incorrectly. The well-organized decision tree can easily categorize a document by putting it in the root node of the tree and let it run through the query structure until it reaches a certain leaf which represents the goal for the classification of the document. The decision tree classification method have several advantages over other decision support tools. The main advantage of
decision tree is that it is easy in understanding and interpreting, even for non-well rounded users. Also, they are robustness to noisy data and they have the ability to learn disjunctive expressions seem suitable for text categorization. The major drawback of
using a decision tree is over fits the training data with the occurrence of an alternative tree that categorizes the
training data worse but would categorize the documents to be categorized better. One of the most well known decision tree algorithm is J-48 that we will use in our work.\\

  \item \emph{K-Nearest Neighbor (KNN):}
  The KNN is one of the simplest lazy classification algorithm \cite{BIB:R21} \cite{BIB:R22} and it is also well known as instance-based learning. The KNN classifier is based on the assumption that the classification of an instance is most similar to the classification of other instances that are nearby in the vector space. To categorize an unknown document d, the KNN classifier ranks the documents among training set and tries to find its k-nearest neighbors, which forms a neighborhood of d. Then majority voting among the categories of documents in the neighborhood is used to decide the class label of d. KNN is an instance-based where the function is only approximated locally and all computation is adjourned until classification. KNN is used in many applications because of its effectiveness, non-parametric and it is easy to be implemented. However, when we use KNN, the classification time is very long and it is not easy to find optimal value of k.  Generally, the best alternative of k to be chosen depends on the data. Also, the effect of noise on the classification is reduced by the larger values of k but make boundaries between the classes less distinct. By using various heuristic techniques, a good 'k' can be selected.
\end{itemize}
\subsection{Performance Measures for Text Classification}
This is the final stage in which the effectiveness of PCA with different text classifier is evaluated. There are different criteria can be used for measuring the performance evaluation of our data sets. In our study we will use the following criteria; confusion matrix, classification accuracy, analysis of sensitivity, specificity and F-measure.
\subsubsection{\emph{Confusion Matrix}} The confusion matrix consist of four classification performance indices: true positive, false positive, false negative, and true negative as given in Table \ref{tab:table1}. They are also usually used in the classification problem to evaluate the performance.
\begin{table}[!htbp]
\renewcommand{\arraystretch}{1.5}
\caption{\small{The four classification performance indices of the confusion matrix.}}
\label{tab:table1}
\centering
\setlength{\tabcolsep}{0.2em}
\begin{tabular}{|c|c|c|}
\hline
\multicolumn{3}{|c|}{Pretdicted Class} \\
    \hline
     {Actual class} & {Classified as \emph{pos}} &{Classified as \emph{neg}} \\
    \hline
     {\emph{pos}}& {True Positive (\emph{TP})} & {False Negative (\emph{FN})} \\
    \hline
    {\emph{neg}}& {False Positive (\emph{FP})} & {True Negative (\emph{TN})}\\
    \hline
\end{tabular}
\end{table}
\subsubsection{\emph{Classification Accuracy}} In our study, the classification accuracy for the data sets are calculated with the following equation:
\begin{equation}\label{e99}
 Accuracy(\%)= \frac{\emph{TP + TN}}{\emph{TP +FN + FP + TN}} \times 100,
\end{equation}
\subsubsection{Analysis of Sensitivity, Specificity and F-measure}
 sensitivity, specificity and F-measure are widely used to evaluate text categorization system. Specificity is the proportion of correctly proposed document to the proposed document. Sensitivity (Recall) is the proportion of the correctly proposed documents to the test data that have to be proposed. For calculating the sensitivity and the specificity of each class we use the following equations.

\begin{equation}\label{e66}
 Sensetivity (\%) = \frac{\emph{TP}}{(\emph{TP+FN})} \times 100,
\end{equation}
\begin{equation}\label{e67}
 Specificity (\%) = \frac{\emph{TN}}{(\emph{TN+FP})} \times 100,
\end{equation}
F-measure is a measure of test's accuracy. It is a harmonic measure of both sensitivity and specificity. It is given by the following equation
\begin{equation}\label{e68}
\small F-measure (\%) = \frac{2 \times \small Sensetivity \times \tiny Specificity}{\small Sensetivity + \small Specificity} \times 100,
\end{equation}

\section{Experimental Results and Analysis}
In this section, we present a series of experiments for text categorization on three different standards and popular text collection to examine the performance of using principle components analysis (PCA) with different text classifiers; standard RF, standard SVM, the J-48 decision tree and KNN methods are used in our study such that they are simple and give acceptable performance measurements in text categorization. The outline of the text data we used is in table \ref{tab:table2}, where the data is downloaded from UCI machine learning databases \cite{BIB:R23}. For the classification stages we select a \emph{10 fold cross validation}. All experiments are run on a personal computer with the configuration of Windows 7 Operation system, 1.8 GHz CPU, 2 GB of RAM and 500 GB HDD space. To evaluate the performance of the different classifiers we use WEKA 3.8.1 software which is developed by the University of Waikato \cite{BIB:R888}.\\
\begin{table}[!htbp]
\caption{\small{Data Set Description}}
\label{tab:table2}
\centering
\setlength{\tabcolsep}{0.2em}
\begin{tabular}{|c|c|c|c|}
\hline
      Data Set & \small No.of Doc & \small NO. of original Features & \small Classes \\
    \hline
 Classic03 & 3830 & 100 & 3 \\
 DBworld & 64 & 229 & 2 \\
   CNAE-9 & 1080 & 856 & 9\\
\hline
\end{tabular}
\end{table}

\begin{table}[!htbp]
\caption{\small{The Performance(average value of precision, sensitivity and F-measure)of RF Classifier on Three Data Sets without PCA}}
\label{tab:table3}
\centering
\setlength{\tabcolsep}{0.2em}
\begin{tabular}{|c|c|c|c|c|c|c|}
\hline
Data set & \small{N. of Features} & Precision  & Sensitivity & F-measure & Accuracy \\
\hline
    Classic03 & 100 & 95.9\% & 95.9\% & 95.9\% & 95.9\% \\

     DBworld &  229 & 82.7\% & 79.7\% & 79.6\% & 79.6\% \\

    CNAE-9 & 856 & 90.4\% & 90.3\% & 90.3\% & 90.2\% \\
\hline
\end{tabular}
\end{table}
In our study, after pre-processing the data by eliminating the stop words which are worthless for classification and using porter algorithm for stemming step \cite{BIB:R13}, we use TF-IDF weighting scheme to reflect the important of the word to the document. This is followed by using PCA, which gives an acceptable results when dealing with text data, to reduce the huge size of the matrix we have. Then finally the extracted features are then passed to various classifiers. We first separately examine the performance of various classifiers, we have mentioned above, without using PCA for feature extraction. Then we examine these various classifiers after using PCA. The aim of using various classifiers is to compare the performance of these methods before and after using PCA.\\\\
\\
Tables \ref{tab:table3}, \ref{tab:table5}, \ref{tab:table7} and \ref{tab:table9} present the results obtained for the three data sets for standard RF, SVM, J-48 and KNN classifiers respectively. However, obtained results in tables \ref{tab:table4}, \ref{tab:table6}, \ref{tab:table8} and \ref{tab:table10} show that using PCA to reduce the features vector before using the standard classifiers improve the accuracy with most of the data sets compared to the standard case.\\
\\

\begin{figure}[!t]
\centering
\includegraphics[width=2.5in]{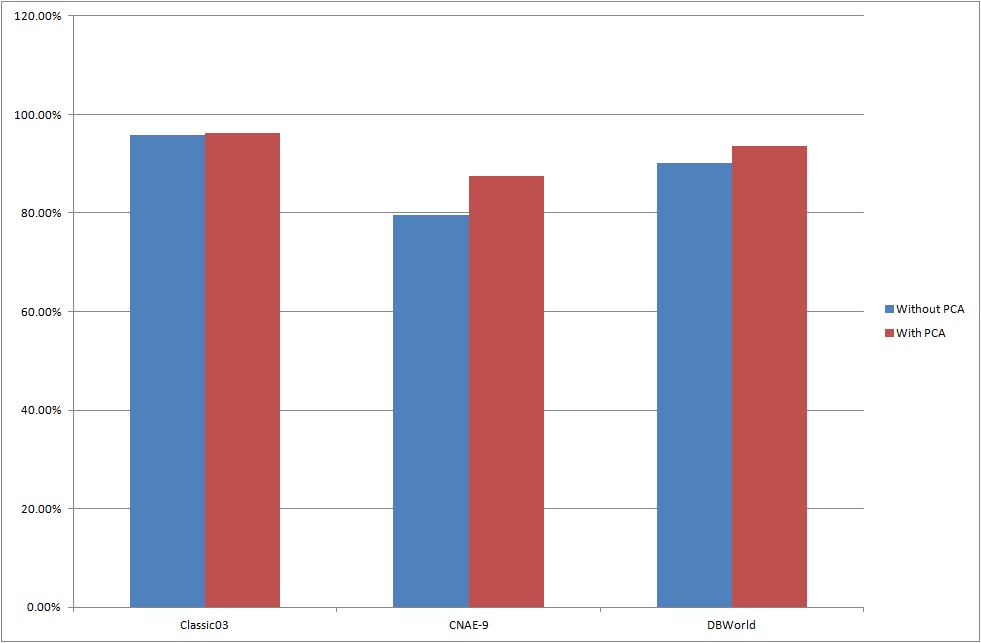}
\caption{Random forest classification accuracy with and without using PCA.}
\label{fig1}
\end{figure}

\begin{figure}[!t]
\centering
\includegraphics[width=2.5in]{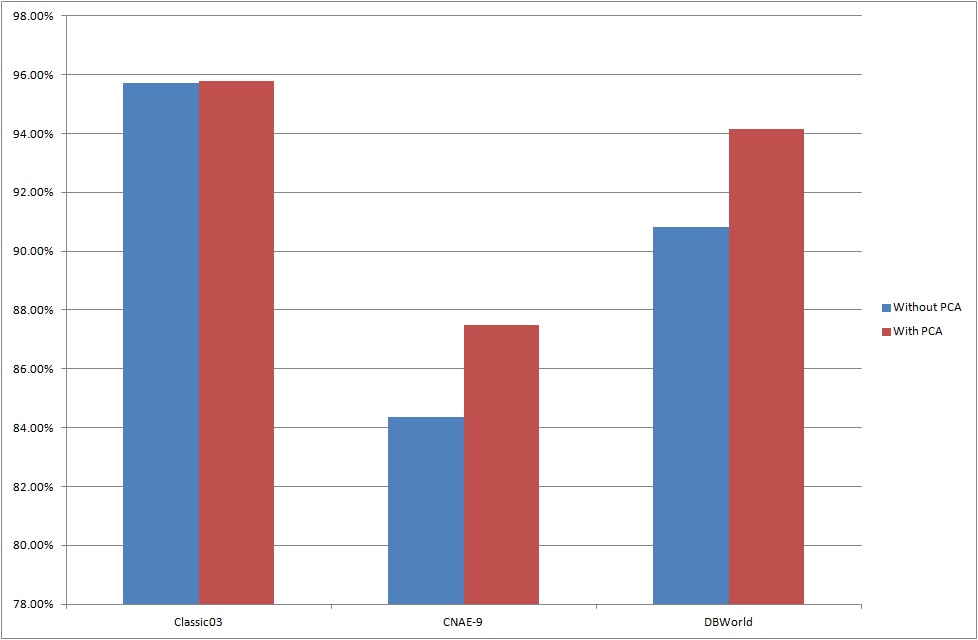}
\caption{Support vector machine classification accuracy with and without using PCA.}
\label{fig2}
\end{figure}

\begin{figure}[!t]
\centering
\includegraphics[width=2.5in]{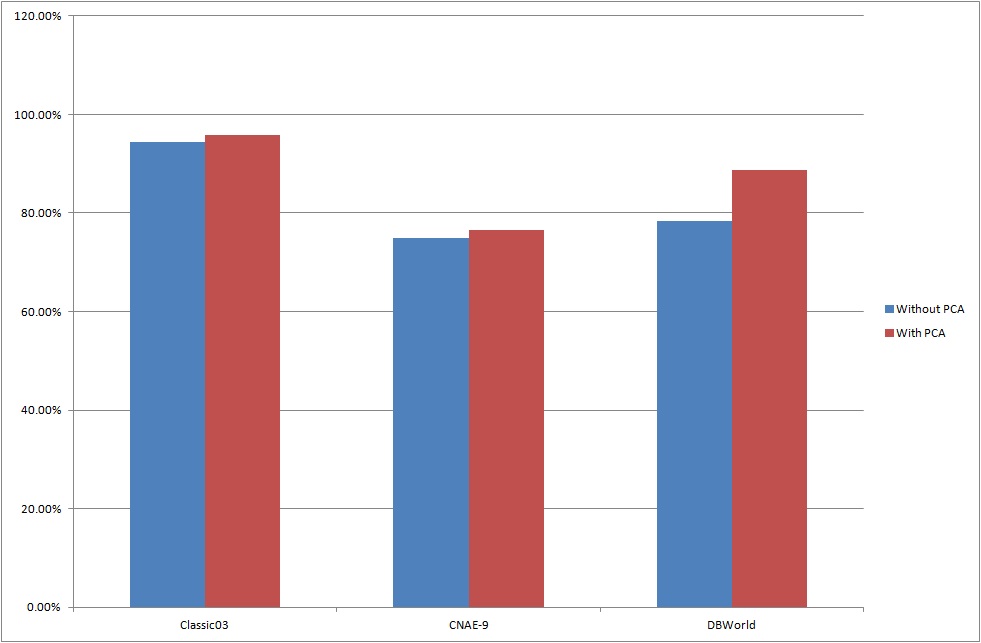}
\caption{ Decision tree (J-48) classification accuracy with and without using PCA.}
\label{fig3}
\end{figure}

\begin{figure}[!t]
\centering
\includegraphics[width=2.5in]{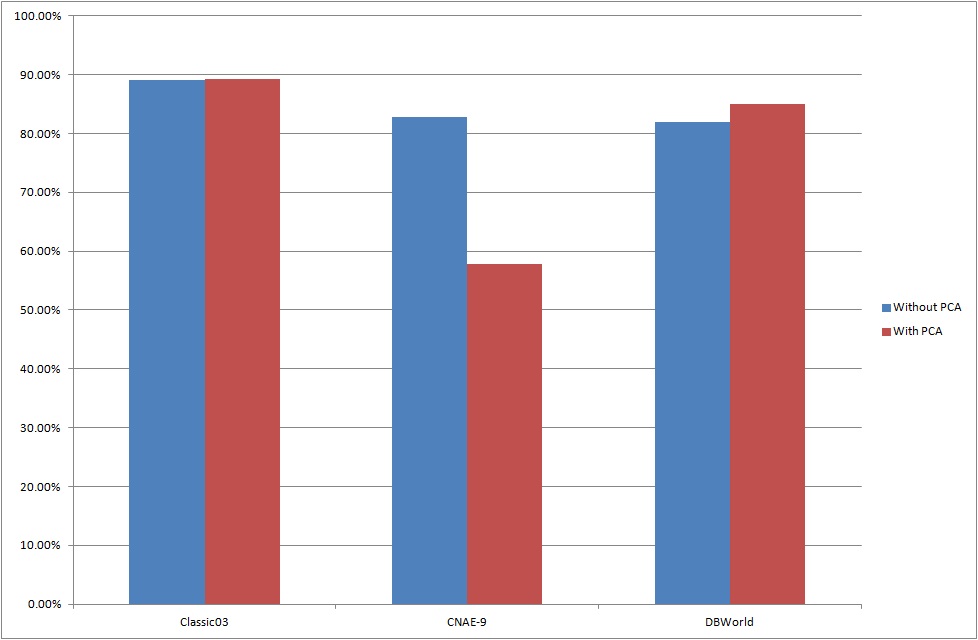}
\caption{K-nearest neighbor classification accuracy with and without using PCA.}
\label{fig4}
\end{figure}

%
%
%
\begin{table}[!htbp]
\caption{\small{The Performance(average value of precision, sensitivity and F-measure)of RF Classifier with PCA}}
\label{tab:table4}
\centering
\setlength{\tabcolsep}{0.2em}
\begin{tabular}{|c|c|c|c|c|c|c}
\hline
Data set & \small{N. of Features} & Precision  & Sensitivity & F-measure & Accuracy \\
\hline
    Classic03 & 86 & 96.4\% & 96.4\% & 96.4\%& 96.3\% \\

     DBworld &  52 & 87.7\% & 87.5\% & 87.5\%& 87.5\% \\

    CNAE-9 & 397 & 93.6\% & 93.6\% & 93.6\% & 93.6\%\\
\hline
\end{tabular}
\end{table}

Results in table \ref{tab:table3} and \ref{tab:table4} show that when using RF as a classifier it gives acceptable and higher accuracy when compared to the rest of classifiers we use. The results in table \ref{tab:table4} show the improvement occur in accuracy after using PCA in the three data sets.\\

\begin{table}[!htbp]
\caption{\small{The Performance(average value of precision, sensitivity and F-measure)of SVM Classifier on Three Data Sets without Using PCA}}
\label{tab:table5}
\centering
\setlength{\tabcolsep}{0.2em}
\begin{tabular}{|c|c|c|c|c|c|c}
\hline
Data set & \small{N. of Features} & Precision  & Sensitivity & F-measure & Accuracy \\
\hline
    Classic03 & 100 & {95.8\%} & {95.7\%} & {95.7\%} & {95.71\%} \\

     DBworld & 229 & 84.4\% & 84.4\% & 84.4\%& 84.37\% \\

    CNAE-9 & 856 & 91.1\% & 90.8\% & 90.9\% & 90.83\% \\
\hline
\end{tabular}
\end{table}
The results in tables \ref{tab:table5} and \ref{tab:table6} show that SVM gives high accuracy with the three data sets and how the accuracy improve after using PCA with SVM.\\\\
\begin{table}[!htbp]
\caption{\small{The Performance(average value of precision, sensitivity and F-measure)of SVM Classifier with PCA}}
\label{tab:table6}
\centering
\setlength{\tabcolsep}{0.2em}
\begin{tabular}{|c|c|c|c|c|c|c}
\hline
Data set & \small{N. of Features} & Precision  & Sensitivity & F-measure & Accuracy \\
\hline
    Classic03 & {86} & {95.9\%} & {95.8\%} & {95.8\%} & {95.79\%}\\

     DBworld & 52 & 87.7\% & 87.5\% & 87.5\% & 87.5\% \\
    CNAE-9 & 397 & 94.4\% & 94.2\% & 94.2\% & 94.16\% \\
\hline
\end{tabular}
\end{table}

 Tables \ref{tab:table7} and \ref{tab:table8} show the performance of J-48 with our text data sets. Combining PCA and J-48 together improve the performance of J-48 with all data sets we use.\\
 Results in tables \ref{tab:table9} and \ref{tab:table10} show the performance of KNN classifier. When we compare the results in the two tables in DBworld data set we find that PCA fail to improve the performance of the KNN classifier.\\
 With the respect to all experimental results shown above, it is seen that RF algorithm gives higher performance with text data when compared with other classifiers and its performance is improved after using PCA.\\
\begin{table}[!htbp]
\caption{\small{The Performance(average value of precision, sensitivity and F-measure)of J-48 Classifier on Three Data Sets without Using PCA}}
\label{tab:table7}
\centering
\setlength{\tabcolsep}{0.2em}
\begin{tabular}{|c|c|c|c|c|c|c}
\hline
Data set & \small{N. of Features} & Precision  & Sensitivity & F-measure & Accuracy \\
\hline
    Classic03 & 100 & 94\% & 94\% & 94\% & 94.4\%\\

     DBworld & 229 & 81.6\% & 75\% & 74.3\% & 75\% \\

    CNAE-9 & 856 & 78.6\% & 78.3\% & 78.4\% & 78.33\% \\
\hline
\end{tabular}
\end{table}
\begin{table}[!htbp]
\caption{\small{The Performance(average value of precision, sensitivity and F-measure)of J-48 Classifier with PCA}}
\label{tab:table8}
\centering
\setlength{\tabcolsep}{0.2em}
\begin{tabular}{|c|c|c|c|c|c|c}
\hline
Data set & \small{N. of Features} & Precision  & Sensitivity & F-measure & Accuracy \\
\hline
    Classic03 &  {86} & {95.9\%} & {95.8\%} & {95.8\%} & {95.79\%} \\

     DBworld & 52 & 79.4\% & 76.6\% & 76.4\% & 76.56\% \\
    CNAE-9 & 397 & 90.4\% & 88.8\% & 89.1\% & 88.79\% \\
\hline
\end{tabular}
\end{table}
\begin{table}[!htbp]
\caption{\small{The Performance(average value of precision, sensitivity and F-measure)of KNN Classifier on Three Data Sets without Using PCA}}
\label{tab:table9}
\centering
\setlength{\tabcolsep}{0.2em}
\begin{tabular}{|c|c|c|c|c|c|c}
\hline
Data set & \small{N. of Features} & Precision  & Sensitivity & F-measure & Accuracy \\
\hline
    Classic03 & 100 & 89.7\% & {89.2\%} & {89.2\%} & {89.1\%} \\

     DBworld & 229 & 86\% & 82.8\% & 82.7\% & 82.81\% \\

    CNAE-9 & 856 & 82.3\% & 81.9\% & 82\% & 81.94\% \\
\hline
\end{tabular}
\end{table}
\begin{table}[ht]
\caption{\small{The Performance(average value of precision, sensitivity and F-measure)of KNN Classifier with PCA}}
\label{tab:table10}
\centering
\setlength{\tabcolsep}{0.2em}
\begin{tabular}{|c|c|c|c|c|c|c}
\hline
Data set & \small{N. of Features} & Precision  & Sensitivity & F-measure & Accuracy \\
\hline
    Classic03 &  {86} & {90\%} & {89.3\%} & {89.3\%} & {89.32\%} \\

     DBworld & 52 & 72.7\% & 57.8\% & 52.4\% & 57.81\% \\
    CNAE-9 & 379 & 85.4\% & 85\% & 84.8\% & 85\% \\
\hline
\end{tabular}

\end{table}
\\

Figures \ref{fig1}, \ref{fig2}, \ref{fig3}, and \ref{fig4} show the comparison of the classification accuracy of various classifiers we use before and after using PCA as a feature extraction technique on the different text data sets. The classification accuracy of most of the classifiers we use is improved after using PCA. Figures show that the improvement performance is much more marked with the classifier random forest (RF) when used on Classic03 data set as compared to other data sets.

\section{Conclusion}
In this study, we use principle component analysis (PCA) as a feature extraction technique to reduce the high dimensionality of the feature vector. PCA removes the irrelevant, noisy and redundant features from the feature vector and thereby improve the performance of the classifier for text categorization. First, we pre-process the documents where the feature vector is obtained through different steps like stop words removal, stemming and indexing. On the core preprocessed documents the classifiers RF, SVM, KNN and J-48 are applied, without dimension reduction, and the performance of the classifiers are observed in terms of sensitivity, specificity and F-measures. Secondly, the feature extraction method principle component analysis (PCA) is applied in the core features and the feature dimension is reduced. Then the classifiers we mentioned above are applied on the extracted features. Most of the obtained results show that the performance of most of the classifiers improved after using PCA and this seems very promising for text categorization applications.\\


\ifCLASSOPTIONcompsoc
  \section*{Acknowledgments}
\else
  \section*{Acknowledgment}
\fi

The authors would like to thank \textbf{Dr. Rasha Mahmoud } for many constructive technical discussions and for comments that greatly improved the manuscript. During my working on this paper, I learnt many things from her including building a system set up, provided guidance for the statistical analysis of my results, etc. I thank her very much for all her help.

\bibliographystyle{ieeetr}
\bibliography{refrence3}


\end{document}